\documentclass[preprint2]{aastex}

\newcommand{\der}{{\rm d}}
\newcommand{\ph}{_{\rm ph}}
\newcommand{\cir}{_{\rm c}}
\newcommand{\ra}{{\cal R}_{\rm a}}
\newcommand{\x}{_{\rm x}}
\newcommand{\mr}{{\cal R}_{\rm m}}
\newcommand{\tang}{_{\rm t}}
\newcommand{\rad}{_{\rm r}}
\newcommand{\rs}{r_{\rm s}}
\newcommand{\maxi}{_{\rm max}}
\newcommand{\modot}{M$_\odot$\ }
\newcommand{\modotc}{M$_\odot$}
\newcommand{\ti}{t_{\rm i}}
\newcommand{\cc}{_{\rm c}}
\newcommand{\delm}{\Delta_{\rm m}}
\newcommand{\delv}{\Delta_{\rm vir}}
\newcommand{\nbody}{{$N$}-body }
\newcommand{\sphc}{(r,\theta,\varphi)}
\newcommand{\beq}{\begin{equation}}
\newcommand{\eeq}{\end{equation}}
\newcommand{\beqa}{\begin{eqnarray}}
\newcommand{\eeqa}{\end{eqnarray}}
\newcommand{\lav}{\langle}
\newcommand{\rav}{\rangle}
\newcommand{\spot}{\lav\Phi\rav}
\newcommand{\srho}{\lav\rho\rav}
\newcommand{\sj}{\lav j\rav}
\newcommand{\jmax}{\sj(R)}
\newcommand{\isph}[1]{\int_0^{2\pi}\!\!\der\varphi\!
            \int_0^{\pi}\!\!\der\theta\;\sin\theta\;{#1}}
\newcommand{\derpr}{\partial_{\rm r}}
\newcommand{\rms}{\sigma_\Phi}

\slugcomment{Submitted to ApJ, February 2007}

\shorttitle{Structure and kinematics DM halos}
\shortauthors{Gonz\'alez-Casado et al.}

\begin{document}

\title{The Accretion-Driven Structure and Kinematics of Relaxed Dark
Halos}

\author{Guillermo Gonz\'alez-Casado}
\affil{Dept.~Matem\`atica Aplicada II, Centre de Recerca d'Aeron\`autica i de l'Espai (UPC-IEEC), \\ Universitat Polit\`ecnica 
de Catalunya, Edifici Omega, Jordi Girona 1--3, E-08034 Barcelona, Spain}
\email{guillermo.gonzalez@upc.edu}

\author{Eduard Salvador-Sol\'e and Alberto Manrique} 
\affil{Departament d'Astronomia i Meteorologia, Institut de Ci\`encies del Cosmos\altaffilmark{1} (UB-IEEC),\\Universitat de
Barcelona, Mart{\'\i} i Franqu\`es 1, E-08028 Barcelona, Spain}

\and

\author{Steen H. Hansen}
\affil{Dark Cosmology Center, Niels Bohr Institute,
University of Copenhagen, \\Juliane Maries Vej 30, 2100 Copenhagen,
Denmark}

\altaffiltext{1}{Associated with the Instituto de
Ciencias del Espacio, Consejo Superior de Investigaciones Cient\'\i
ficas.}


\begin{abstract}
It has recently been shown that relaxed spherically symmetric dark
matter halos develop from the inside out, by permanently adapting
their inner structure to the boundary conditions imposed by the
current accretion rate.  Such a growth allows one to infer the typical
density profiles of halos. Here we follow the same approach to infer
the typical spherically averaged profiles of the main structural and
kinematic properties of triaxial, anisotropic, rotating
halos. Specifically, we derive their density, spatial velocity
dispersion, phase-space density, anisotropy and specific angular
momentum profiles. The results obtained are in agreement with
available data on these profiles from \nbody simulations.
\end{abstract}



\section{INTRODUCTION}\label{intro}

High-resolution cosmological simulations show that relaxed cold dark
matter (CDM) halos are close to triaxial homologous systems (Bailin \&
Steinmetz 2005) with a variety of axial ratios but essentially
universal profiles. That is, the shape of the spherically averaged
radial profile of any given structural and kinematic property is
always very similar, independently of halo mass, epoch, environment and
cosmology considered; only the scaling may depend on such
particularities.

Dubinski \& Carlberg (1991) and Crone et al.~(1994) noted that halos
of very different masses show similar scaled density profiles. Navarro
et al.~(1997, hereafter NFW) then showed that the spherically averaged
density profile is always well fit, down to about one hundredth the
virial radius $R$, by the simple expression
\beq
\lav\rho\rav(r)\propto \frac{1}{r(\rs+r)^2}\,,\label{rho}
\eeq
with the scale radius $\rs$ correlating with the total halo mass $M$
within $R$ in such a way that the smaller $M$, the higher the
concentration $c\equiv R/\rs$, a correlation that was interpreted
(NFW; Salvador-Sol\'e et al.~1998) as due to the fact that less
massive halos typically form earlier when the mean cosmic density is
higher.

Although there is nowadays general agreement on the previous universal
density profile, some authors claim that higher resolutions yield
slightly steeper central cusps (Fukushige \& Makino 1997, 2001; Moore
et al.~1998, 1999; Ghigna et al.~2000; Jing \& Suto 2000; Diemand et
al.~2004), while others advocate rather the opposite, that the density
profile becomes shallower as smaller and smaller radii are reached
(Taylor \& Navarro 2001; Power et al.~2003; Fukushige et al.~2004;
Hayashi et al.~2004). Zhao (1996) has proposed a more general
practical expression that accounts for all these
possibilities. Besides, it has more recently been shown (Navarro et
al.~2004; Merritt et al.~2005; Merritt et al.~2006; Graham et al.~2006;
see also Stoehr et al.~2002) that a new expression of the S\'ersic
(1968), in 3D, or Einasto (1965) law
\begin{equation}
\srho(r)=\rho_0\exp\left[-\left(\frac{r}{r_{\rm n}}\right)^{1/n}\right]\,,
\label{new}
\end{equation}
with strictly no central cusp, yields still better fits to the
mass distribution of simulated halos. Down to the
resolution-limited radii reached by current simulations, the
discrepancies among all these analytical fits are of the order of the
deviates found using any individual one (Dehnen \& McLaughlin 2005;
see also Navarro et al.~2004 and Fukushige et al.~2004), which
explains such a diversity of opinions. In respect to the $c(M)$
relation, different analytical expressions have also been proposed
that try to recover the results of numerical simulations at various
redshifts (NFW; Eke, Navarro \& Steinmetz 2001; Bullock et al.~2001a;
Zhao et al.~2003). Again, there is overall agreement between them
although the discrepancies become substantial as one gets apart from
the mass and redshift ranges analyzed in simulations.

On the other hand, the 3D velocity dispersion profile $\sigma(r)$ is
well fit by the solution of the Jeans equation for spherical,
isotropic systems resulting from the empirical NFW density profile and
vanishing velocity dispersion at infinity (Cole \& Lacey 1996; see
also Merritt et al. 2006 for the Einasto profile). What is more
interesting, as noted by Taylor \& Navarro (2001; see also Ascasibar
et al.~2004; Rasia et al.~2004; Barnes et al.~2006), the (pseudo)
phase-space density profile, $\lav\rho\rav/\sigma^3$, is always close
to a pure power law in radius
\beq
\frac{\lav\rho\rav(r)}{\sigma^3(r)}= A r^{-\nu}\,,\label{tn}
\eeq
with index $\nu\approx 1.9$ (see also Dehnen \& McLaughlin 2005 for a
similar relation applying to the radial component of the velocity
dispersion). 

In addition, Hansen and Moore (2006) have recently shown that the
pressure-supported aniso\-tro\-py profile, 
\beq \beta(r) \equiv
1-\frac{\sigma\tang^2(r)}{\sigma\rad^2(r)}=\frac{1}{2}\left[3-\frac{\sigma^2(r)}
{\sigma^2\rad(r)}\right]\,,
\label{beta}
\eeq
defined as usual in terms of the 1D radial and tangential velocity
dispersion profiles, $\sigma\rad(r)$ and $\sigma\tang(r)$,
respectively, is related to the logarithmic slope of the spherically
averaged density profile through the simple linear relation
\beq
\beta(r)=a \left(\frac{\der \ln \lav\rho\rav}{\der \ln r}+b\right)\,,
\label{hm}
\eeq
with constants $a$ and $b$ respectively equal to $\approx -0.2$
and $\approx 0.8$ (Hansen \& Stadel 2006).

The simplicity of the relations (\ref{tn}) and (\ref{hm}) might
suggest that they are more fundamental than the universal halo density
and velocity dispersion profiles themselves. However, they appear to
be equivalent: not only do the latter profiles satisfy the previous
equations, but they are also the only profiles to do so. Indeed, the NFW
density profile is the only physically acceptable, realistic (with no
central hole), profile with Zhao's (1996) general form that solves
the Jeans equation satisfying the relation (\ref{tn}), both in the
simple isotropic case (Austin et al.~2005; see also Taylor \& Navarro
2001 and Hansen 2004) and the general anisotropic one, provided in
this latter case the additional relation (\ref{hm}) (Dehnen \& McLaughlin
2005).

Finally, cosmological simulations show that relaxed CDM halos have a
small angular momentum, typically orientated along the minor axis of
the inertia ellipsoid, whose modulus $J$ is log-normally
distributed (e.g., Barnes \& Efstathiou 1987; Ryden 1988; Warren et
al.~1992; Cole \& Lacey 1996; Bullock et al.~2001b, hereafter B01b;
Bailin \& Steinmetz 2005) in terms of the dimensionless spin parameters,
\beq
\lambda=\frac{J|E|^{1/2}}{GM^{5/2}}\,,
\label{lambda}
\eeq
or 
\beq
\lambda'=\frac{J}{\sqrt{2} M R V\cir}=\frac{J}{M\sqrt{2GMR}}\,,
\label{lambdap}
\eeq
respectively defined by Peebles (1980) and B01b, where $G$ is the
gravitational constant, $E$ is the total energy of the halo (with
vanishing potential at infinity), $R$ is the virial radius and
$V\cir\equiv \sqrt{GM/R}$ is the circular velocity. According to their
respective definitions, both spin parameters would coincide provided
halos were singular isothermal spheres, but for halos endowed with the
NFW density profile, one has (B01b)
\beq
\lambda'\approx \lambda F(c)\,,
\label{lamlamp}
\eeq
with 
\beq
F^{-2}(c)=\frac{2}{3}+\left(\frac{c}{21}\right)^{0.7}\,.
\label{fc}
\eeq
The mean $\lambda'$ value appears to be essentially constant in time
(Hetznecker \& Burkert 2006) but dependent on halo mass, while the
mean $\lambda$ value is independent of mass (B01b) but shows a slight
variation in time (Hetznecker \& Burkert 2006) due, in principle, to
the evolution of halo concentration and mass distribution.

The local specific angular momentum vector within halos tends to be
everywhere aligned (B01b), the cumulative mass-distribution of
specific angular momenta $M(<j)$ being well fit by a simple
two-parameter function and the spherically averaged specific angular
momentum profile $\sj(r)$ fairly well fit by the simple expression
\beq
\sj(r)=\jmax\,\left[\frac{M(r)}{M}\right]^s\,,
\label{jr}
\eeq
where $M(r)$ is the mass inside $r$ and index $s$ takes values around 1.3.

As argued by Austin et al.~(2005), the equilibrium condition alone
cannot explain all these universal trends because the Jeans equation
is not restrictive enough, nor can the initial conditions, as very
different protohalos lead to essentially the same final structural
(Austin et al.~2005; Romano-Diaz et al.~2006) and kinematic (Hansen \&
Moore 2006) properties. Thus, what can only be at their origin is the
way these systems grow and, in the case of the specific angular
momentum profile, the effects of tidal torques suffered during that
process (Doroshkevich 1970; White 1984).

Two extreme points of view have been investigated. Some authors have
looked at the possibility that the universal density profile arises
essentially from the effects of repeated mergers (Syer \& White 1998;
Salvador-Sol\'e et al.~1998; Subramanian et al.~2000; Dekel et
al.~2003). Others have focused on smooth accretion through spherical
infall (Avila-Reese et al.~1998; Nusser \& Sheth
1999; del Popolo et al.~2000; Manrique et al. 2003; Ascasibar et
al.~2004). Similarly, the origin of the universal specific angular
momentum profile has been studied in the spherical collapse
approximation (B01b) or by considering the cumulative effect of
mergers (Gardner 2001; Maller et al.~2002; Vitvitska et al.~2002).

A notable result found along this latter line of research is that the
density profile predicted from spherical accretion resembles greatly
that found in numerical simulations. However, major mergers cannot be
ignored in any realistic hierarchical cosmology, which gives little
credit to the ``pure'' accretion scenario. Yet, the density profiles
of halos do not depend on the epoch they suffered the last major
merger (Wechsler et al.~2002) nor, in general, on their particular
aggregation history (Huss et al.~1999; Romano-Diaz et
al.~2006). Manrique et al.~(2003) pointed out that this would be well
understood if the inner structure of relaxed halos were completely
fixed by the boundary conditions imposed by {\em current} accretion
and did not depend on the halo past history. In fact, taking into
account that halos develop from the inside out during accretion phases
(Salvador-Sol\'e et al.~2005; Romano-Diaz et al.~2006; Lu et
al.~2006), their typical density profile can be readily derived, in
the spherically symmetric case, from the accretion rate characteristic
of the cosmology under consideration (Manrique et al.~2003).

This accretion-driven density profile is in very good agreement with
the results of numerical simulations in the whole radial, mass and
redshift ranges reached by them (Salvador-Sol\'e et al.~2007,
hereafter SMGH) and recovers all the correlations shown by simulated
halos such as the mass-concentration relation (Salvador-Sol\'e et
al.~2005). Furthermore, as recently shown by SMGH, all the conditions
required in the derivation of this profile, but the spherical symmetry
assumed for simplicity, emanate directly from the very nature of
standard CDM. The fact that CDM is collisionless guarantees that the
spatial distribution of any physical quantity is continuous at any
derivative order. As a consequence, all halos with mass $M$ at the
time $t$ undergoing the same accretion $\dot M(t)$ during some finite
time interval around $t$ have identical (non-scaled) $M(r)$ and,
hence, $\rho(r)$ profiles regardless of their individual past
history. On the other hand, the non-decaying, non-self-annihilating
and dissipationless nature of CDM added to the fact that it accretes
slowly onto halos guarantees their inside-out growth. Finally, the CDM
power-spectrum characteristic of the particular cosmology considered
fixes the typical accretion rate and through it the typical density
profile of relaxed halos of any given mass at any given cosmic time.

Yet, this success would be of limited interest if the same explanation
did not also hold for the mass distribution in more realistic,
triaxial halos and for any other property apart from the density. In
the present paper, we show that this approach allows one to understand
all the universal structural and kinematic trends of triaxial,
velocity-anisotropic, rotating halos. In Section \ref{uni}, we show
that the total values at a given cosmic time of the extensive
quantities characterizing relaxed halos and the rates at which they
increase during any finite time interval around that moment determine
completely the corresponding spherically averaged profiles. This is
used, in Section \ref{prof}, to infer the typical shape of these
profiles. Our results are summarized and discussed in Section
\ref{dis}. Throughout the present paper we adopt the concordance
cosmological model characterized by $(\Omega_{\rm
m},\Omega_\Lambda,h,\sigma_8)=(0.3,0.7,0.7,0.9)$.

\section{ACCRETION RATES AND UNIQUENESS OF PROFILES}\label{uni}

Consider a relaxed (i.e. quasi-steady), triaxial,
velocity-anisotropic, rotating dark matter halo. The mass inside
radius $r$ takes the usual form for spherical systems,
\beq
M(r)=4\pi \int_0^r \der\tilde{r}\;\tilde{r}^2\; \lav\rho\rav (\tilde{r})\,,
\label{sphm}
\eeq
in terms of the spherically averaged density profile, 
\beq
\srho(r) = \frac{1}{4\pi} \isph{\rho\sphc}\,,
\label{avden}
\eeq
where $\rho\sphc$ is the local density in spherical coordinates $r$,
$\theta$ and $\varphi$, centered at the peak density and with $z$ axis
orientated along the total angular momentum {\bf J} within $R$.

The kinetic energy inside $r$ also takes the form for spherical
systems (see the detailed derivation in the Appendix)
\beq K(r)=4\pi \int_0^{r} \der \tilde r\,\tilde r^2\srho(\tilde r)
\frac{\sigma^2(\tilde r)}{2}\,,
\label{energyr}
\eeq
in terms of the usual velocity-non-centered spatial velocity dispersion
profile, $\sigma(r)$, which coincides with the spherical average of
the local velocity dispersion $\sigma\sphc$ in the $r$
shell\footnote{In contrast, the {\it velocity-centered} velocity
dispersion at $r$, defined as the rms velocity deviation {\it from the
here non-vanishing mean value}, differs in general from the spherical
average of the corresponding local quantity.}. Note that in dealing
with {\it non-centered} velocity dispersions, the influence of rotation
on the total kinetic energy of the system is included only implicitly
in equation (\ref{energyr}).

Simulations show that the local angular momentum vector is essentially
aligned all over the halo, eventually except for the innermost and
outermost regions respectively affected by the limited resolution and
possible boundary effects due to the inclusion of infalling matter
(B01b; Bailin \& Steinmetz 2005). Consequently, the modulus $J$ of the
total angular momentum vector inside $r$ can be written in terms of
the modulus $j$ of the local specific one as
\beq
J(r)=\!\int_0^r\!\!\der\tilde r\,\tilde r^2\!\!\!
\int_0^{2\pi}\!\!\der\varphi\!
            \int_0^{\pi}\!\!\der\theta 
\sin\!\theta\,j\sphc\,\rho\sphc\,.
\label{am1}
\eeq

Thus, using the spherical average profile,
\beq
\sj(r)\!=\!\frac{1}{4\pi\srho(r)}
\!\!\int_0^{2\pi}\!\!\!\!\der\varphi\!\!
            \int_0^{\pi}\!\!\der\theta 
\sin\!\theta j\sphc \rho\sphc,
\label{sj}
\eeq
it takes again the same form as for spherically symmetric systems,
\beq J(r) = 4\pi
\int_0^r\!\der r\;r^2\,\sj(r)\,\srho(r)\,.
\label{am2}
\eeq

To conclude this preamble, let us add that the similarity with the
spherically symmetric case when dealing with spherically symmetric
profiles does not stop here. It also concerns, although only {\it
approximately}, other profiles such as the potential energy inside
$r$ or relations such as the Jeans equation and the scalar virial
relation (see the Appendix). In particular, from equations
(\ref{pressure}) and (\ref{virapr}) one has the relation
\beq 
\sigma\rad^2(R)\! =\! \frac{1}{ R^3 \srho (R)}\!
\int_0^{R}\!\!\! \der r\, r^2\srho(r)
\left[\sigma^2(r)\!-\!\frac{GM(r)} {r}\right]\!.
\label{virialR}
\eeq

\subsection{Density Profile}\label{dens}

As mentioned, relaxed halos evolve inside-out between major mergers,
that is, they keep their inner density distribution unaltered and just
stretch it outwards (Salvador-Sol\'e et al.~2005; Romano-Diaz et
al.~2006; Lu et al.~2006). This is the natural consequence of the fact
that centered spheres of arbitrary radii conserve the mass (standard
CDM is non-decaying and non-self-annihilating) and energy (standard
CDM is dissipationless) and that the characteristic accretion time of
halos is substantially smaller than their dynamical time (see SMGH). It
is true that relaxed halos are triaxial, so they can in principle
suffer tidal torques from the surrounding matter affecting the kinetic
energy of such inner spheres. However, the tidal field produced by the
surrounding anisotropic large scale mass distribution is very stable
(possibly except for the short time interval before major mergers) and
relaxed halos remain elongated along one fixed privileged
direction. Consequently, tidal torques have a minimal effect on the
internal kinematics of relaxed halos; they are only important near
maximum expansion of protohalos where the angular momentum of those
structures is generated.

The inside-out growth implies that, at any time $t$ during such
accretion periods, the total mass is given by (see eq.~(\ref{sphm}])
\beq
M(t)\equiv M[R(t)]=4\pi\int_0^{R(t)} \der r\,r^2\srho(r)\,,
\label{mass}
\eeq
where the function $\srho$ inside the integral on the right is
independent of time. Thus, by differentiating eq.~(\ref{mass}) we are
led to
\beq
\srho(t)\equiv \srho[R(t)]=\frac{\dot M}{4\pi R^2(t)\dot R}\,.
\label{rhot}
\eeq
As the virial radius encompasses a region with inner mean density equal
to some factor $\delv(t)$ (e.g. Bryan and Norman 1998) times the mean
cosmic density $\bar\rho(t)$,
\beq
R(t)=\left[\frac{3M(t)}{4\pi\delv(t)\bar\rho(t)} \right]^{1/3}\,,
\label{rvir}
\eeq
$\dot R$ in equation (\ref{rhot}) is a function of $\dot
M$. Then, all relaxed halos with the same value of $M$ (and $R$)
at $t$, accreting mass at the same rate $\dot M$ during any finite
time interval around $t$, will develop identical $\srho$ profiles over the
corresponding finite radial range. As CDM is collisionless and
free-streaming and, hence, it admits no discontinuity in the steady
spatial distribution of any structural or kinematic property, the
function $\srho(r)$ must be analytical. Consequently, if the density
profiles of such halos coincide over some finite radial range, they
necessarily do at {\em any other radius.}  As explained in SMGH, this means
that the density profile of relaxed halos permanently adapts to {\em
current accretion} and, hence, does not depend on the halo's past
aggregation history.

\subsection{Velocity Dispersion Profile}\label{disper}

As relaxed halos are in equilibrium, the fact that their mass
distribution develops inside-out automatically implies that their
local velocity tensor keeps unaltered as they accrete. Indeed, centered
spheres of any arbitrary radius conserve not only the mass but also
the kinetic energy (and the gravitational energy as well provided the
origin of the potential remains unchanged; see the discussion
below). Consequently, all kinematic profiles, in particular the
velocity dispersion profile, must develop inside-out just like the
spherically averaged density profile. Therefore, the total energy is
given by (see eq.~[\ref{etot}])
\beqa
E(t)\equiv E[R(t)]=4\pi \int_0^{R(t)} \der r\,r^2\srho(r) & \nonumber\\
\times\,\left[\frac{\sigma^2(r)}{2}-\frac{GM(r)}{r}\right]\,,&
\label{energy}
\eeqa
with all the functions within the integral on the right independent of
time and, by differentiation, we obtain
\beq 
\sigma^2(t)\equiv \sigma^2[R(t)] =2\left[\frac{\dot E}{\dot M} +
\frac{GM(t)}{R(t)}\right]\,.
\label{sig2}
\eeq
Caution must be paid to the fact that the two preceding expressions
presume the gravitational potential, with origin at infinity, for the
system truncated at the virial radius. This simplifies notably the
expression of the total energy as it depends on the inner mass
distribution only, which is particularly useful when dealing with
protohalos at very early times (see below). Had we not adopted that
point of view, we would have been led to the more general expression
\beq 
\sigma^2(t)\equiv \sigma^2[R(t)] = 2\frac{\dot E}{\dot M} +
\spot[R(t)]
\label{sig2bis}
\eeq
where $\spot(r)$ is the spherically averaged gravitational potential
(eq.~[\ref{spot}]) associated to the in general non-truncated mass
distribution.

In any case, the same reasoning above leading to the uniqueness of the
$\srho$ profile for halos with the same value of $M$ at $t$ accreting
mass at the same rate $\dot M$ during some finite time interval around
$t$ leads now to the uniqueness of their $\sigma$ profile if they also
accrete energy at the same rate $\dot E$ during that interval. In
other words, the (analytical) $\sigma$ profile also permanently adapts
to the energy accretion currently undergone by the halo.

\subsection{Anisotropy Profile}\label{anis}

According to the preceding discussion, the $\srho$ and $\sigma$
profiles of a relaxed halo are completely set by the {\it
independent\/} mass and energy accretions it is currently
undergoing. What enables the emerging structure to be in equilibrium
is the freedom provided by the anisotropy, which adapts to produce a
steady configuration consistent with the inside-out growth. More
specifically, for some given $\srho$ and $\sigma$ profiles, the
(approximate) Jeans equation (\ref{exJeq2}) for anisotropic systems is
a differential equation for $\sigma\rad(r)$, which fixes the
anisotropy profile.  To see it consider the virial relation
(\ref{virialR}) resulting from integration of the Jeans
equation. Taking into account the relation (\ref{energy}), we can
write
\beqa 
\sigma\rad^2(t)\equiv \sigma\rad^2[R(t)]= \left[4\pi R^3(t)
\,\srho (t)\right]^{-1}~~~ &\nonumber\\
\times\,\left[ 2E(t)+\int_0^{R(t)} \der M(r)\frac{GM(r)}{r}\right]\,.&
\label{sigrad}
\eeqa
Thus, the value of the radial velocity dispersion at any given radius
is completely determined indeed by both independent $\srho$ and
$\sigma$ profiles down to the halo center. 

To sum up, the fact that relaxed halos develop inside-out implies that
those having identical values of $M$ and $E$ at $t$ and undergoing the
respective accretions at the same rates $\dot M$ and $\dot E$ during
any finite time interval around $t$ necessarily have not only
identical $\srho$ and $\sigma$ profiles, but also identical
$\sigma\rad$ (and, hence, $\sigma\tang$ as well) and $\beta$ profiles.

\subsection{Angular Momentum Profile}
\label{sam}

As mentioned, the expression (\ref{energyr}) for the kinetic energy
within $r$ includes the influence of rotation. As the mass and kinetic
energy of halos in centered spheres of arbitrary radii are kept
unaltered during accretion, the total angular momentum inside them
must also necessarily be kept unchanged. The total angular momentum of
accreting halos may, of course, change with time. But, according to
the preceding reasoning, this can only be done at the expense of the
matter newly incorporated at the instantaneous radius like for the
total mass and energy of the system.

Under these circumstances, equation (\ref{am2}) leads to
\beq 
J(t) \equiv J[R(t)]= 4\pi \int_0^{R(t)}\!\!\der
r\,r^2\sj(r)\srho(r)\,,
\label{am2t}
\eeq
where all the functions inside the integral on the right are
again independent of time and, by differentiating, we obtain the relation
\beq \sj(t)\equiv \sj[R(t)] =\frac{\dot J}{\dot M}\,.
\label{jt}
\eeq

Then, the same reasoning leading to the uniqueness of the $\srho$,
$\sigma$ and $\beta$ profiles, added now to the approximation of
aligned local angular momenta, leads to the fact that halos with given
values of $M$ and $J$ (or of $M$, $E$ and $\lambda$ or $\lambda'$) at
$t$, increasing at the same respective rates $\dot M$, and $\dot J$
(or $\dot M$, $\dot E$ and $\dot\lambda$ or $\dot\lambda'$) during any
finite time interval around $t$, also have identical $\sj$ profile at
all radii. 

\begin{figure}
\vspace{-70pt}
\centerline{
\includegraphics[width=0.80\textwidth]{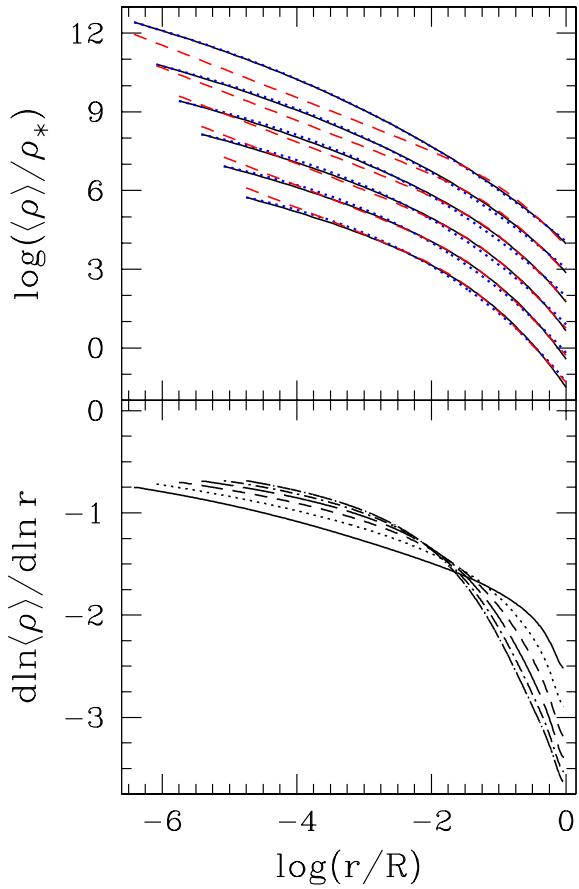}
}
\caption{Predicted density profiles (top panel) as a function of
radial distance in units of virial radius, R, for halos at $z=0$
(black solid lines), compared to their best fit by a NFW law (red
dashed lines) and by a S\'ersic law (blue dotted lines) down to R/100
and 1 pc, respectively, for halos with $10^{10}$, $10^{11}$,
$10^{12}$, $10^{13}$, $10^{14}$ and $10^{15}$ \modot (from bottom to
top). The normalization constant $\rho_\ast$ is defined as $\bar\rho
M_\ast/M$ where $M_\ast=10^{13}$ \modot is the critical mass for
collapse. Corresponding radial dependence of logarithmic slopes (bottom
panel) for the same halo masses: $10^{10}$ \modot (dot-long dashed
line), $10^{11}$ \modot (dot-short dashed line), $10^{12}$ \modot
(long dashed line), $10^{13}$ \modot (short dashed line), $10^{14}$
\modot (dotted line) and $10^{15}$ \modot (solid line).}
\label{1}
\end{figure}

\section{TYPICAL PROFILES}\label{prof}

The fact that all the preceding profiles are unique for given current
values of the corresponding accretion rates is not of much help, in
general, for inferring them for any particular halo. This requires
performing the analytical extension of the respective small pieces
developed by accretion in any finite time interval, which is not an
easy task. However, what can be readily derived in the way explained
next is the {\em typical} radial behavior of all these profiles.

\subsection{Density Profile}\label{dens2}

As noted by SMGH, the analytical extension of the typical mass profile
$M(r)$ is simply the transformation from $t$ to $r$ through the
inverse of equation (\ref{rvir}) of the analytical $M(t)$ track solution 
of the differential equation
\beq
\frac{\dot M}{M(t)}=\ra[M(t),t]\,,
\label{acrate}
\eeq
where 
\beq
\ra[M(t),t]= \int_0^{M \delm}\!\!\! \der(\Delta M)\;
\Delta M\;\mr(M,t,\Delta M)\,
\label{track}
\eeq
is the analytical typical scaled accretion rate in the cosmology under
consideration (Raig et al.~2001). Then, one must simply differentiate
such a mass profile to obtain the desired typical density profile,
\beqa
\srho(t)=\delv(t)\bar\rho(t)~~~~~~~~~~~~~~~~~~~~~~~~~~~~~ &\nonumber\\
\times\,\left[1-\frac{1}{\ra[M(t),t]}\;
\frac{\der\ln(\delv\bar\rho)}{\der t}\right]^{-1}\,,&
\label{rhor}
\eeqa
for $t(r)$ equal to the inverse of $R(t)$ given by equation
(\ref{rvir}). In equation (\ref{track}), $\mr(M,t,\Delta M)$ is the
usual merger rate in the extended Press-Schechter (PS) formalism
(Lacey \& Cole 1993) and the integral extends only over those mergers
producing a relative mass increase below the effective ratio
$\delm=\Delta M/M$ separating minor from major mergers as the former are
the only ones to contribute accretion. As shown in SMGH, the best
effective value of this threshold, equal to 0.26, gives an excellent
fit to the empirical NFW $c(M)$ relation over at least 4 decades in
halo mass (see Fig. 2 in SMGH).

\begin{figure*}
\vspace{-8cm}
\centerline{\includegraphics[width=0.85\textwidth]
{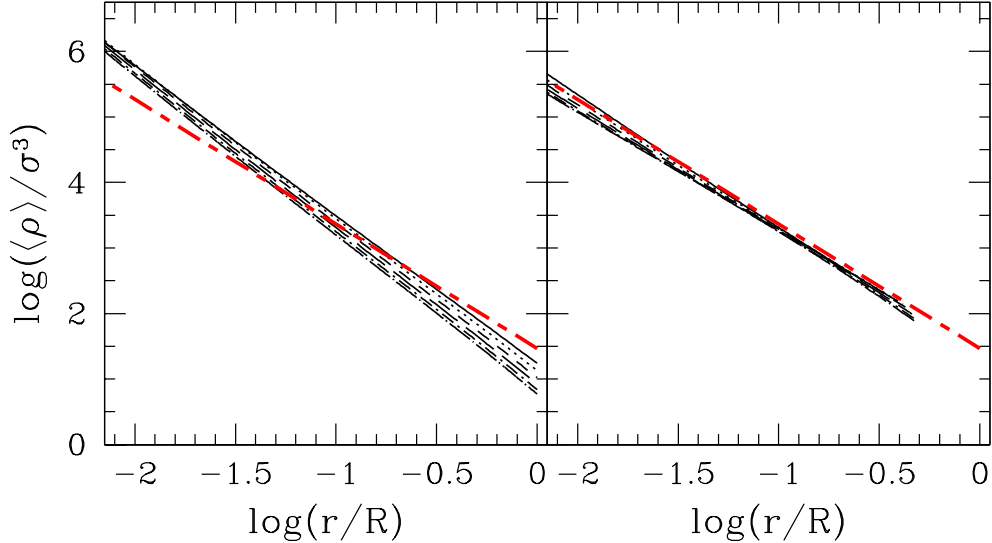}}
\caption{Predicted phase-space density profiles at $z=0$ for the same
halo masses as in Figure \ref{1} (and same symbols as in the bottom
panel of that figure). The red thick long dashed-short dashed line is the
empirical curve obtained by Ascasibar et al.~(2004). Left panel:
predictions obtained from the spherical collapse model without
shell-crossing. Right panel: predictions obtained from the
phenomenological correction for shell-crossing with $\eta=1.8$.
The density and velocity dispersion are in units of the cosmic
critical density and of halo circular velocity, respectively.}
\label{2}
\end{figure*}

\begin{figure}[t]
\vspace{-30pt}
\centerline{\includegraphics[width=0.59\textwidth]{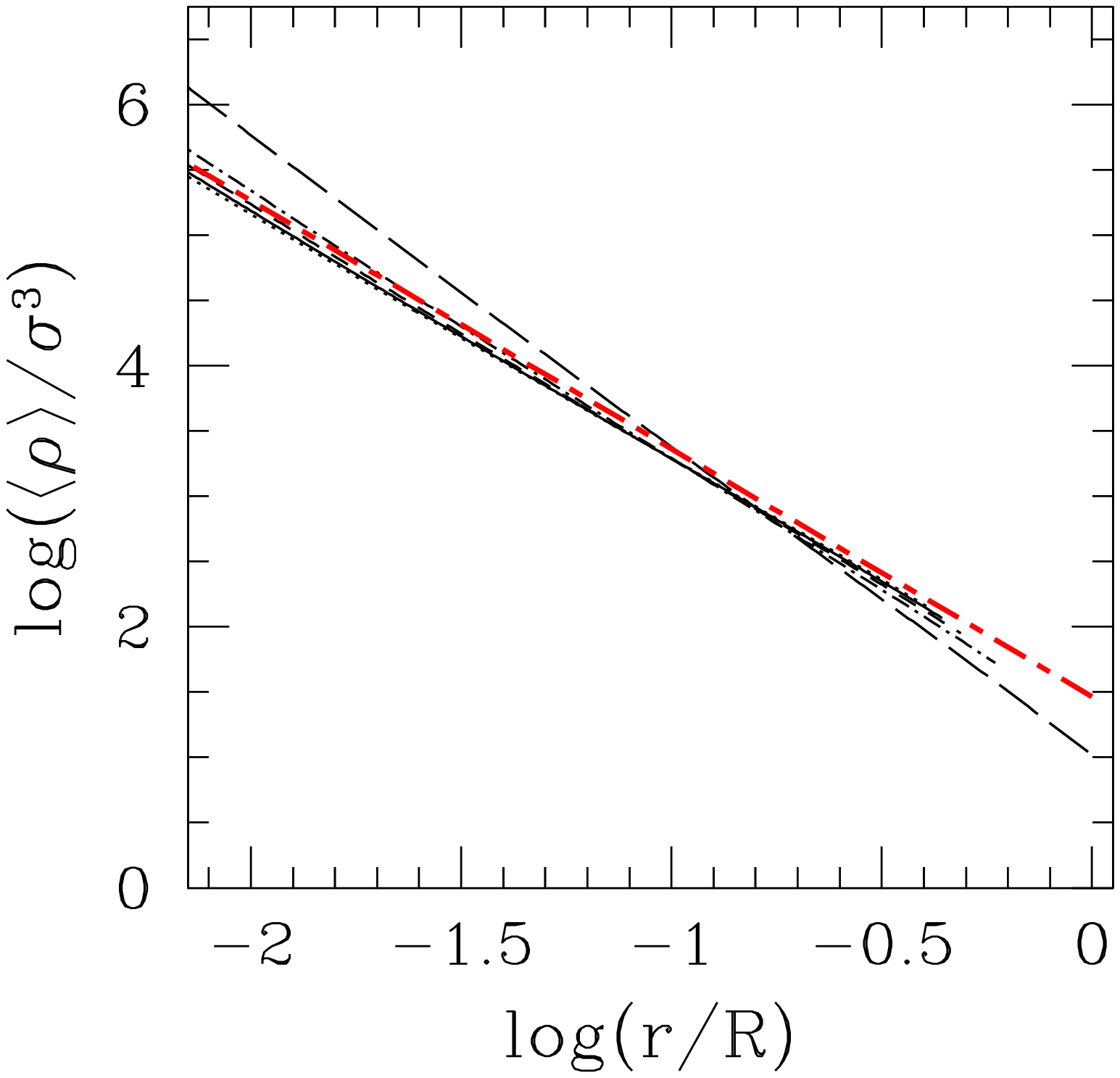}}
\vspace{-15pt}
\caption{Predicted phase-space density profile for a halo of $10^{13}$
\modot at $z=0$ for different values of $\eta$: $0$ (long dashed
line), $1$ (dot-dashed line) $1.5$ (short dashed line), $1.8$
(solid line) and $2$ (dotted line) compared to the empirical curve
obtained by Ascasibar et al.~(2004) (red thick long dashed-short dashed
line).}
\label{3}
\end{figure}

In the top panel of Figure \ref{1} we show the predicted density
profiles, for $\delm=0.26$, corresponding to various halo masses at
$z=0$, compared to their respective best fits down to a radius of one
hundredth $R$ and 1 pc by a NFW law (eq.~[\ref{rho}]) and an Einasto
law (eq.~[\ref{new}]), respectively. As mentioned, not only is the
density profile well fit by an NFW law down to one hundredth the virial
radius as found in numerical simulations, but the mass-concentration
relation is also perfectly recovered. In respect to the fit by an
Einasto law, there is so far no accurate mass dependence of the
corresponding parameters in the literature; see SMGH for that
predicted by the present model. In the bottom panel of Figure \ref{1}
we plot the corresponding logarithmic slopes as a function of radius,
where the tendency to approach, at small radii, the functionality
$\der \ln \srho/\der \ln r\propto -r^{1/n}$ characteristic of the
Einasto law is more apparent.

\subsection{Velocity Dispersion Profile}
\label{disper2}

To derive the typical $\sigma$ profile we need to know the typical
energy accretion rate $\dot E$ going along with the typical mass
accretion rate $\dot M$ given above. Although there is no expression
similar to (\ref{acrate}) for $\dot E$, we can try to estimate it by means
of conservation arguments from the typical total energy of the halo seeds
at any arbitrarily small time $\ti$.

When the total mass $M(t)$ of virialized objects at $t$ is to be
estimated it is very useful to approximate protohalos by smooth
spherical top-hat perturbations. This does not work however when
dealing with the total energy $E(t)$ because it depends on the halo
mass and velocity distributions at all scales. (Nor does it help
taking into account that protohalos would coincide with peaks of the
{\it smoothed} density field.) Determining accurately the total energy
of the protohalo is, in general, a very difficult task. But, in the
present case, we can take advantage of the fact that the structure and
kinematics of the final halo do not depend on its past aggregation
history and adopt the point of view that it has evolved since $\ti$ by
pure accretion. That assumption and the neglect of triaxiality effects
as done in the Appendix for the final relaxed halos\footnote{Such {\it
accreting\/} protohalos are also much centrally peaked, which
guarantees the validity of the approximation.}  allow one to deal with
those seeds as if they were spherically symmetric.

In this case, the radius $R\ph$ of that part of the protohalo
collapsing at $t$ is
\beq
R\ph^3(t)=\frac{3M(t)}{4\pi\bar\rho(\ti)[1+\delta(t,\ti)]}\,,
\label{rtophat}
\eeq
where $\delta(t,\ti)=\delta\cc D(\ti)/D(t)$ is the density contrast
for spherical collapse at $t$, with $\delta\cc$ the critical value for
current collapse (equal to 1.69 in any flat cosmology) and $D(t)$ the
perturbation linear growth factor. Consequently, its mass and total
energy (for the system truncated at $R\ph$ and with potential origin
at infinity) are
\beq
M(t)=4\pi\int^{R\ph(t)}_0 \der r\; r^2\;\rho\ph(r)
\label{mtophat}
\eeq 
\beq
E(t)=4\pi\!\!\int^{R\ph(t)}_0 \!\!\der r\, r^2\rho\ph(r)\!
\left[\frac{u^2\ph(r)}{2}\!-\!\frac{GM\ph(r)}{r}\right]\!\!,
\label{etophat}
\eeq 
where $\rho\ph(r)$ and $M\ph(r)$ are the exact, spherically averaged,
density and mass profiles, respectively, of the protohalo and
$u\ph(r)$ is the velocity of the shell at $r$, essentially equal to
the Hubble component, $H(\ti)r$. Differentiating over $R\ph(t)$ both
$M(t)$ and $E(t)$, respectively given by equations (\ref{mtophat}) and
(\ref{etophat}), and substituting these derivatives into equation
(\ref{sig2}), one is led, after some algebra and to leading order in
the perturbation $\delta(t,\ti)$, to
\beq \sigma^2(t)= 2\frac{GM(t)}{R(t)}
\left[1-\frac{\delta\cc}{\delv^{1/3}(t)}\right]\,,
\label{sigma2t}
\eeq 
yielding, for $t$ equal to the inverse of $R(t)$, the wanted typical
$\sigma^2(r)$ profile. Equation (\ref{sigma2t}) coincides with the
expression we would have obtained from the usual top-hat
approximation, so the detailed density and velocity distributions in
the protohalo do not actually play any role in the final result. The
reason for this is that the specific density profile $\rho\ph(r)$
cancels when taking the ratio between the time derivatives of $E$ and
$M$ in equation (\ref{sig2}). This is the consequence of the fact
that, as the velocity dispersion profile is developing inside-out, its
value at $r$ depends only on the specific energy of the protohalo at
the corresponding radius, not on its total inner value.

The phase-space density profile $\srho/\sigma^3$ arising from such a
velocity dispersion profile and the spherically averaged density
profile derived in Section \ref{dens2} is shown in the left panel of
Figure \ref{2}. For comparison, we plot the empirical best-fitting
power law of index $\nu=1.9$ obtained by Ascasibar et al.~(2004). The
theoretical prediction agrees with the result of numerical simulations
(despite the lack of any free parameter to adjust) not only in its
overall shape, close to a power law over more than five decades in
halo mass, but also in its magnitude. The only small discrepancy is
that the predicted profile is somewhat steeper than empirically found
($\nu\sim 2.3$ instead of $1.9$). This is most likely due to the
fact that we have neglected shell-crossing.

Shell-crossing brakes the infall of collapsing matter, so the bulk
velocity of deep enough layers is much smaller than predicted by the
top-hat collapse model without shell-crossing. In fact, the system is
essentially steady within the virial radius $R$ (Cole \& Lacey
1996). Unfortunately, shell-crossing cannot be dealt with
analytically, so the mass within $R(t)$ can only be estimated
phenomenologically from the predictions of that simple collapse
model. Next a similar approach is used to estimate the total energy of
the object within $R(t)$.

The mass within the shell of the protohalo at $R\ph$ having an inner
mean density contrast appropriate for collapse, without
shell-crossing, at $t$ appears to coincide with the mass $M$ of the
final steady object within the virial radius $R$, defined through
equation (\ref{rvir}), at that moment. This does not mean, of course,
that the particles inside $R$ coincide with those initially located
inside $R\ph$. Some have rebound and are currently beyond $R(t)$,
while others initially beyond $R\ph$ have already passed by
$R(t)$. Yet, as far as the mass within $R(t)$ coincides with that of
the protohalo within $R\ph$, the mass of these two kinds of particles
balance each other and we must not worry about that
distinction. However, the situation is different when dealing with
total energy. Particles having bounced were originally more tightly
bound than those having not yet. Thus, the total energy within $R(t)$
is larger than estimated through equation (\ref{etophat}) (kinetic
energy transfer among shells through two-body interactions at shell
crossing is negligible). In other words, it is given by that equation
at a larger time and the same result holds for the velocity dispersion
at $R$: it is given by equation (\ref{sigma2t}) at some later time
$t+\Delta t$. Inspired of the fact that the shift $\Delta t$ in the
case of $M(t)$ is null for all halo masses and cosmologies, a
reasonable guess for the shift $\Delta t$ in the case of energies is
to assume it equal to $\eta t$ with $\eta$ equal to the same positive
constant factor for all halo masses and cosmologies. This leads to
(see eq.~[\ref{sigma2t}])
\beq \sigma^2(t)= 2\frac{GM[(1+\eta)t]}{R[(1+\eta)t]}
\left\{1-\frac{\delta\cc}{\delv^{1/3}[(1+\eta)t]}\right\}\,.
\label{sigma2tcor}
\eeq 
Moreover, since the shift $\eta t$ is caused by shell-crossing and the
characteristic bounce time of shells that would collapse at $t$ if
there were no shell-crossing is equal to $t$, we expect an $\eta$
value of order unity.

\begin{figure*}
\centerline{\includegraphics[bb=18 144 592 450,clip,width=0.85\textwidth]
{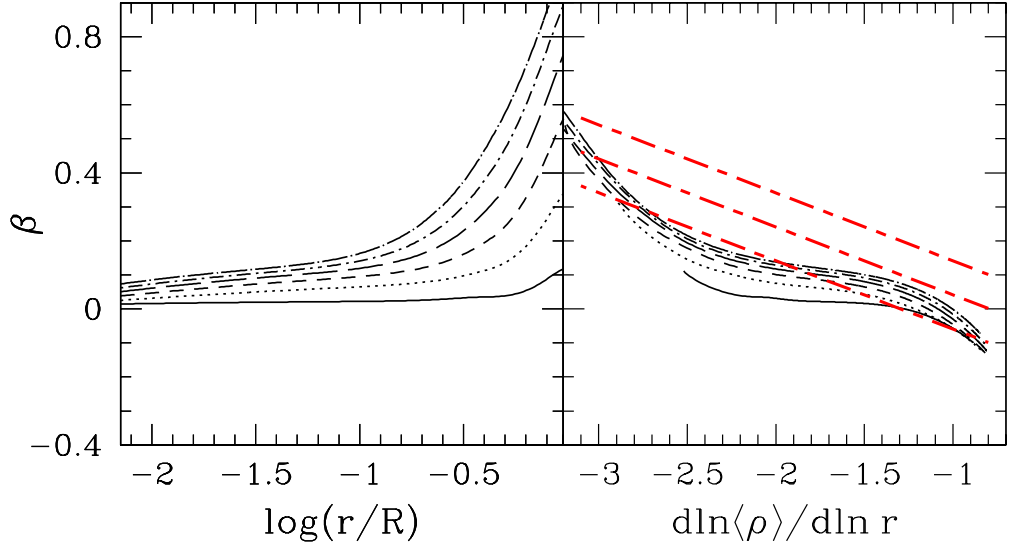}}
\caption{Predicted anisotropy profile as a function of radius (left
panel) or density logarithmic slope (right panel) obtained for the
same halo masses as in Figure \ref{2}. For comparison, we plot the best
linear fit to empirical data on this latter relation drawn from
numerical simulations (central red thick short dashed-long dashed line) and
its scatter (bracketing red thick lines). For $r$ tending to zero, the
predicted anisotropy tends to become negative. Whether this is a real
effect or a numerical artifact (caused, e.g., by the extrapolation of
our phenomenological correction for shell crossing down to $r=0$) is
hard to tell at this stage.}
\label{4}
\end{figure*}

\begin{figure*}
\centerline{\includegraphics[bb=18 144 592 450,clip,width=0.85\textwidth]
{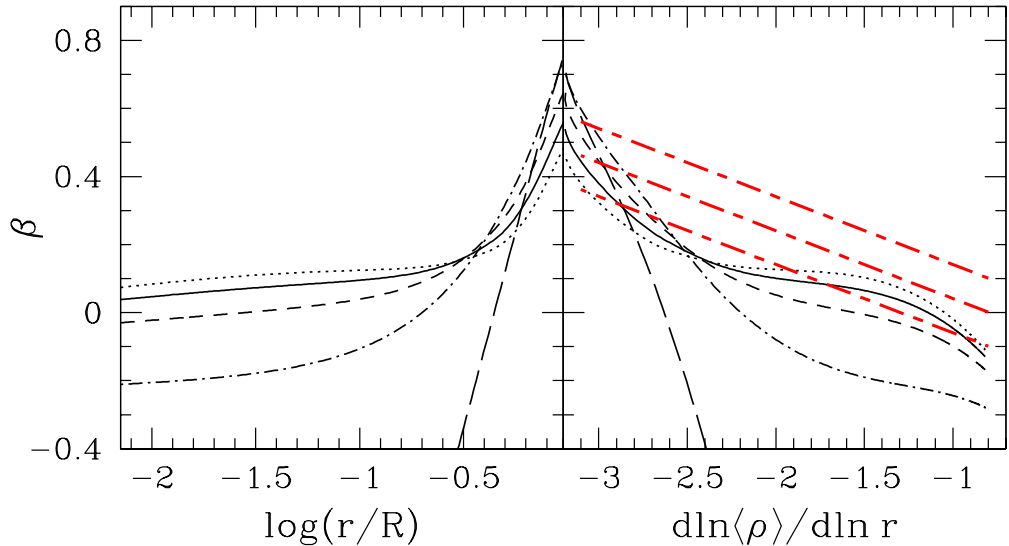}}
\caption{Same as Figure \ref{3} for the predicted anisotropy profile,
showing the effect of the phenomenological factor $\eta$.  Full line
($\eta = 1.8$) gave a good fit when comparing the theoretical
$\srho/\sigma^3$ to those of simulations, and it is also seen to provide
a reasonable fit to $\beta$.}
\label{5}
\end{figure*}
In the right panel of Figure \ref{2}, we depict the phase-space
density profile arising from the spatial velocity dispersion given in
equation (\ref{sigma2tcor}) for $t(r)$ equal to the inverse of $R(t)$
and $\eta=1.8$. As can be seen, this new $\sigma$ profile is in much
better agreement, indeed, with the results of simulations. 

To see the effect of varying $\eta$, we show, in Figure \ref{3}, the
profiles resulting from different values of that parameter. For any
value of $\eta$ greater then zero, the solution $\sigma(r)$ at large
enough radii relies on the extrapolation towards the future of the
Bryan and Norman (1998) expression for $\delv$. For $\eta=1.8$, this
affects the solutions at radii above $\log (r/R)\sim -0.35$ for all
halo masses (see Figs.~\ref{2} and \ref{3} where all profiles are
plotted only for smaller radii). As such an extrapolation is quite
uncertain, from now on, the corrected $\sigma$ profile at such large
radii is inferred from the much more secure linear extrapolation of the
predicted phase-space density profile.

\begin{figure}
\vspace{-30pt}
\centerline{
\includegraphics[width=0.85\textwidth]{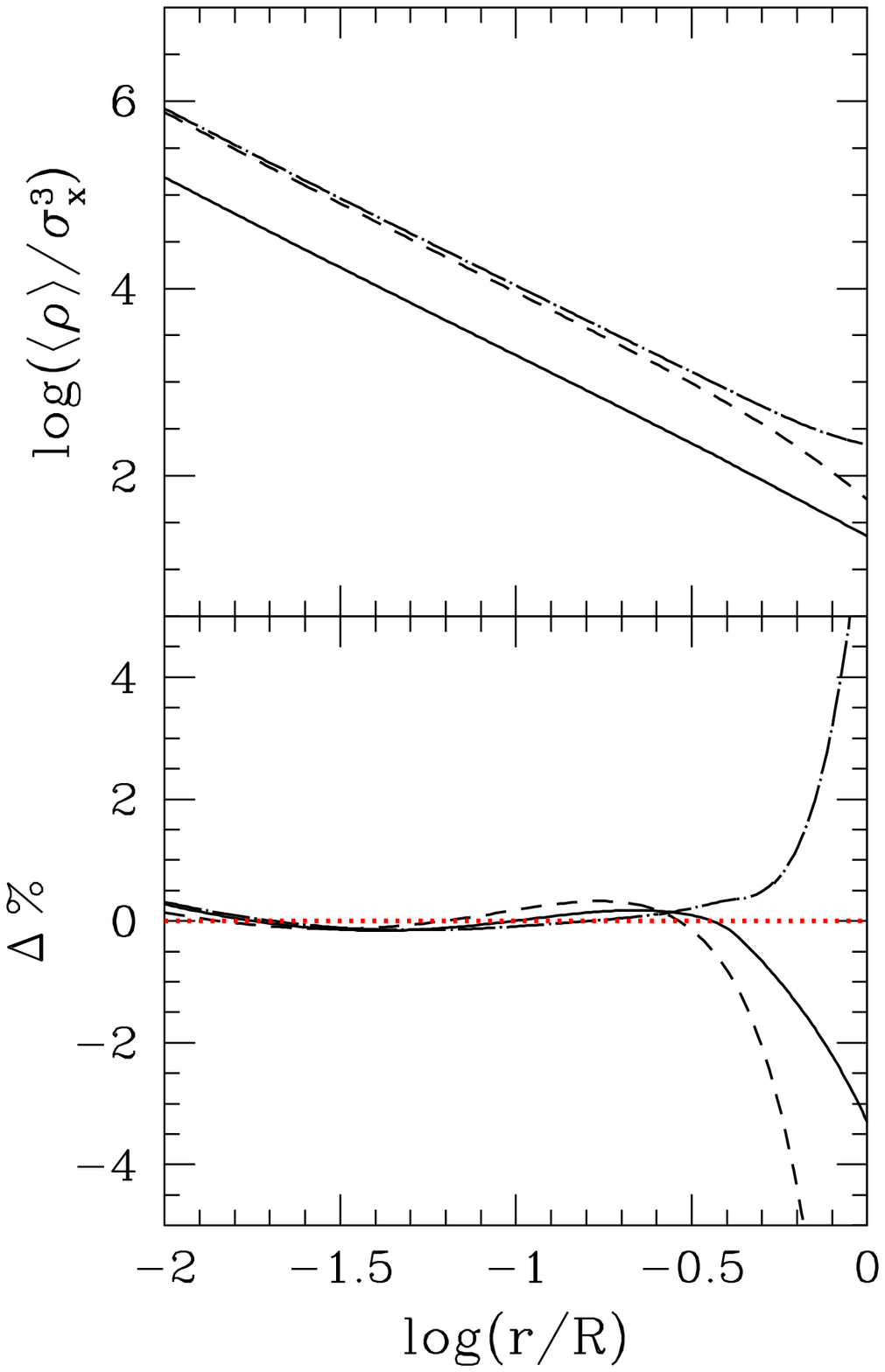}}
\caption{Top panel: predicted phase-space densities for a halo of
$10^{13}$ \modot at $z=0$ corresponding to the radial ($\sigma\x\equiv
\sigma\rad$, dashed line), tangential ($\sigma\x\equiv \sigma\tang$,
dot-dashed line) and spatial ($\sigma\x\equiv \sigma$, solid line)
velocity dispersions obtained with $\eta=1.8$ as
in the right panel of Figure \ref{2}. Bottom panel: residuals from the
respective fit to a power-law of the form $Ar^{-\nu}$ with best-fitting
exponents equal to $1.89$ (spatial case), $1.92$ (radial case) and
$1.87$ (tangential case); all fits achieved in the radial range
requiring no extrapolation.}
\label{6}
\end{figure}

\begin{figure}
\vspace{-100pt}
\centerline{
\includegraphics[width=0.85\textwidth]{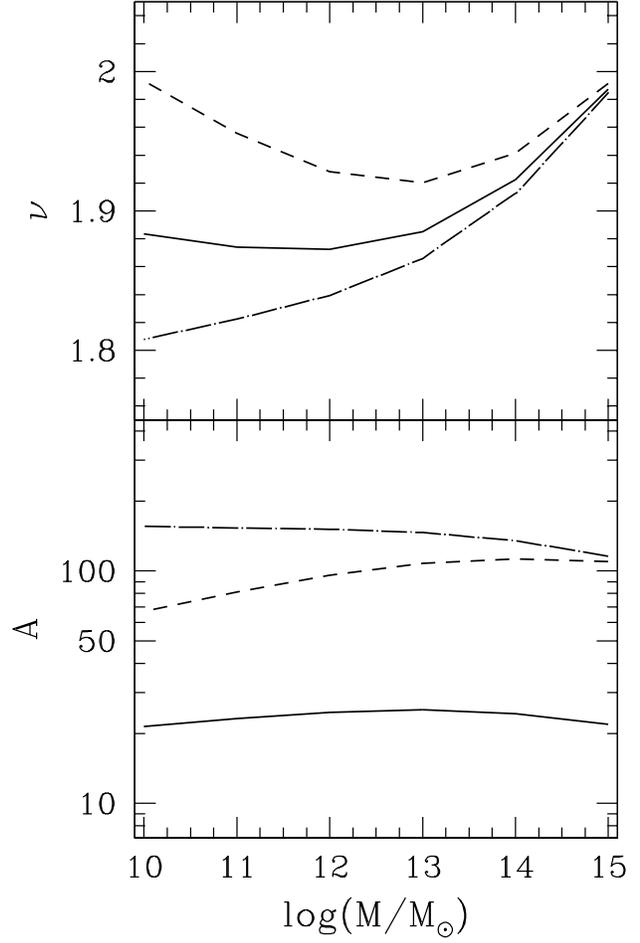}}
\caption{Mass dependence of the power index $\nu$ (top panel) and the
proportionality factor $A$ (bottom panel) of the best-fitting power
law $Ar^{-\nu}$ for the predicted spatial, radial and tangential
phase-space densities (same symbols as in Fig. \ref{6}).}
\label{7}
\end{figure}

\subsection{Anisotropy Profile}\label{anis2}

After a change of integration variable, the approximate relation
(\ref{sigrad}) takes the form
\beqa 
\sigma^2\rad(t) = \frac{1}{4\pi R^3(t)\,\srho (t)}
~~~~~~~~~~~~~~~~~~~~~~~~~~~&\nonumber\\
\!\times \!\!\int_0^t \!\!\der \tilde t \left[\sigma^2(\tilde t)-\frac{GM(\tilde t)}{R(\tilde t)}
\right]\! \ra[M(\tilde t),\tilde t]\,M(\tilde t),~&
\label{sigmar}
\eeqa
leading, for $t(r)$ equal to the inverse of $R(t)$, to the
(approximate) typical radial velocity dispersion profile and, through
equation (\ref{beta}), to the (approximate) typical $\beta(r)$
profile.  Note that, as $\sigma$ was derived to leading order in the
perturbation $\delta$ at $\ti$, its central behavior is not fully
reliable, so is not either that of the $\sigma\rad$ profile. However, the
behavior at moderate and large radii of $\sigma\rad(r)$ is quite
robust because the integral appearing in equation (\ref{sigmar}) is
very insensitive to the central values of $\sigma$.

In the left panel of Figure \ref{4}, we plot the predicted typical
$\beta(r)$ profile for the same current halos and the same radii as in
Figure \ref{2}. In the right panel, we show the corresponding relation
between $\beta$ and the logarithmic slope of the density profile. The
theoretical anisotropy shows a trend similar to that empirically found
by Hansen \& Moore (2006), although with some apparent
undulations. Similar variations have recently been observed in
numerical experiments (McMillan et al.~2007). Whether
these undulations are real or an artifact due to the approximate
character of our solution is hard to tell. Note that the rough
universality of the $\beta$ profile combined with the fact that the
typical halo density profile at $t$ depends only on the halo mass
leads to the conclusion that the typical anisotropy profile depends
essentially only on the halo mass, too.

Of course, such a predicted typical anisotropy profile depends on the
exact value of $\eta$ chosen to correct the theoretical velocity
dispersion profile for the effects of shell-crossing. To see more
quantitatively the influence of such a correction, we plot in Figure
\ref{5} the solutions arising from the different values of $\eta$ used
in Figure \ref{3}.

Once we have determined the typical radial and tangential velocity
dispersion profiles we can check if the corresponding phase-space
density profiles are again close to power laws. The result is shown in
Figure \ref{6} where we also plot the residuals of each profile from
the corresponding best fit by a power-law in $r$ (eq.~[\ref{tn}]). As
can be seen, the phase-space density profiles for the radial and
tangential velocity dispersions also admit a power law fit like that
associated with the spatial velocity dispersion with similar values of
index $\nu$ and just a small shift in the respective proportionality
factors, in agreement with the results of numerical simulations
(Ascasibar et al 2004; Dehnen \& McLaughlin 2005). Among these two
profiles that associated with the tangential dispersion,
$\srho/\sigma^3\tang$, is the one giving the best fit to a power-law
in radius (in the range requiring no extrapolation), with a slope
$\nu\sim 1.9$. This result is most likely related to the fact that the
tangential velocity distribution function has virtually the same shape
for all radii (Hansen et al.~2006). This is contrasted with the radial
velocity distribution function, whose shape changes significantly as
function of radius.

In Figure \ref{7} we show the mass dependence of the best fitting
parameters $\nu$ and $A$ to all three phase-space density profiles.

\subsection{Angular Momentum Profile}\label{sam2}

Equations (\ref{jt}) and (\ref{acrate}) define, for $t(r)$ equal to
the inverse of $R(t)$, the $\sj(r)$ profile in terms of $\dot J$,
which can be computed from equation (\ref{lambda}) or (\ref{lambdap}),
assuming typical values for both the spin parameter and its time
derivative.

As mentioned, the mean spin values and standard deviations are found
in \nbody simulations to be quite insensitive to the cosmology, halo
mass and particular epoch considered. When looking at their behavior
in more detail, it is observed however that, just after major mergers,
they take values substantially higher than the mean (Burkert \&
D'Onghia 2004), which is likely due to the fact that halos are not yet
fully relaxed. What is more important for our purposes here,
$\lambda'$ depends on halo mass according to equation (\ref{lamlamp})
(B01b), its mean value being essentially constant while the mean
$\lambda$ value is essentially independent of mass (B01b) and shows a
slight secular evolution (Hetznecker and Burkert 2006). This latter
result is found regardless of whether the means are performed over all
the halos or just accreting ones. According to the relation
(\ref{lamlamp}), the spin parameters of a given accreting halo satisfy
the relation
\beq 
\lambda'(t) = \lambda(t) F\{c[M(t),t]\}\,.
\label{lp}
\eeq 
Taking the mean of this latter equation over halo masses in the
relevant mass range at $t$ and taking into account that the mean
$\lambda'$ is constant and equal to 0.039 \footnote{Strictly, this
value corresponds to the mean for all halos, that for accreting halos
being slightly smaller (Hetznecker and Burkert 2006). However, this is
a very small effect that can be neglected at this stage.}, one is led
to
\beq 
\lambda(t) = \frac{0.039}{\bar F(t)}\,.
\label{Lambda}
\eeq 
where $\bar F(t)$ stands for the mean of $F[c(M,t)]$ with $F(c)$ given
by equation (\ref{fc}) and $c(M,t)$ the NFW mass-concentration
relation at $t$ given in SMGH (see their eq.~[10]). Note that, at
present, $c$ is around $10$ for masses in the relevant range (between
$10^{10}$ \modot and $10^{15}$ \modotc), so $\bar F$ is about
unity. We have checked that the time-dependence of the mean $\lambda$
value implied by the approximate relations (\ref{lp})--(\ref{Lambda})
is also a reasonable approximation of the one found by Hetznecker and
Burkert (2006).

\begin{figure*}[t]
\centerline{\includegraphics[bb=18 144 592 450,clip,width=0.85\textwidth]
{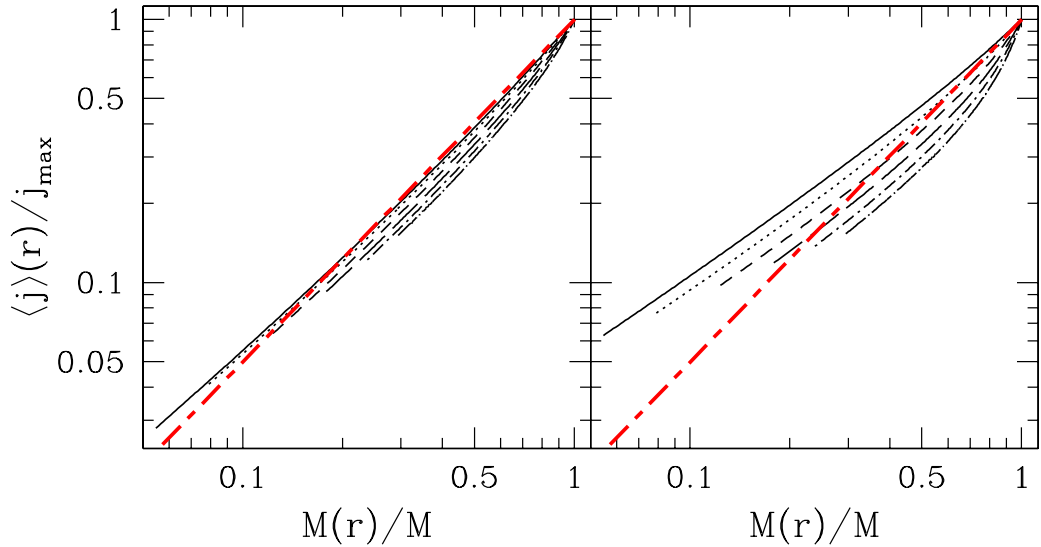}}
\caption{Predicted specific angular momentum profile for the same halo
masses (marked with identical symbols) as in Figures \ref{2} and
\ref{4} derived using the $\lambda$ (left panel) and $\lambda'$ (right
panel) spin parameters. For comparison, we plot the average power law
with $s=1.3$ fitting, to a first approximation, numerical data
according to B01b (red thick short dashed-long dashed line).}
\label{8}
\end{figure*}

\begin{figure*}
\centerline{\includegraphics[bb=18 144 592 450,clip,width=0.85\textwidth]
{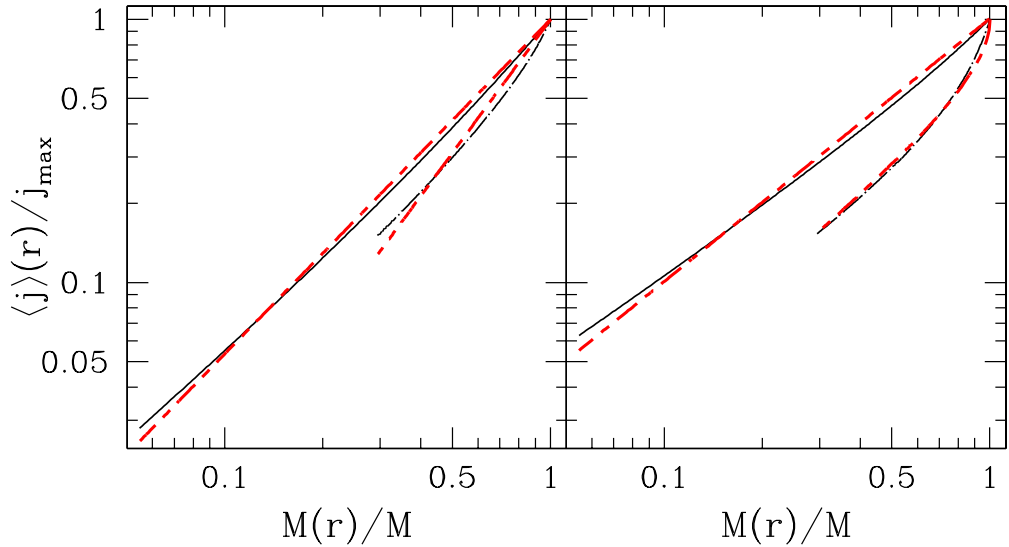}}
\caption{Same predicted specific angular momentum profiles (and same
symbols) as in Figure \ref{8}, and the corresponding fits either by a
power law for the $\lambda$ case (left panel), or by the slight modification of
it given in the text for the $\lambda'$ case (right panel). To avoid overcrowding, we
only represent the curves corresponding to $10^{10}$ \modot and
$10^{15}$ \modotc.}
\label{10}
\end{figure*}

Differentiating equations (\ref{lambda}) and (\ref{lambdap}), we
obtain the typical angular momentum accretion rate
\beq 
\frac{\dot J}{J(t)} = \frac{\dot M}{M(t)}
\Biggl\{
\frac{5}{2}-\frac{1}{\ra[M(t),t]}
\left[\frac{\dot E}{2E(t)}-\frac{\dot\lambda}{\lambda(t)}\right]
\Biggr\}
\label{dotJ}
\eeq
and
\beqa
\frac{\dot J}{J(t)} = \frac{\dot M}{M(t)}
\Biggl\{\frac{5}{3}-\frac{1}{\ra[M(t),t]}~~~~~~~~~~~~~~~~~ &\nonumber\\
\times\,
\left[\frac{\der\ln(\delv
\bar\rho)^{1/6}}{\der t} -\frac{\dot
\lambda'}{\lambda'(t)}\right] \Biggr\}\,,&
\label{dotJp}
\eeqa
respectively. In equation (\ref{dotJ}), the energy accretion rate
$\dot E$ is computed from equation (\ref{sig2bis}), with $\sigma(t)$
according to the expression (\ref{sigma2tcor}) and $\spot$ the
spherically averaged gravitational potential at $R$ of the halo
endowed with a NFW profile up to infinity, and $E(t)$ is obtained by 
integrating $\dot E$ over time. In deriving equation (\ref{dotJp}) from
equation (\ref{lambdap}), we have taken into account the definition of
the virial radius (eq.~[\ref{rvir}]) and the expression of $\srho(t)$
given by equation (\ref{rhor}). Note that the expression (\ref{dotJp})
making use of $\lambda'$ does not depend neither on the
phenomenological correction for shell-crossing of the $\sigma$ profile
nor on the assumed mass distribution beyond $R(t)$.

By substituting into equation (\ref{jt}) the previous expressions for
$\dot J$ and taking into account the typical values of the spin
parameters given by expressions (\ref{lp}) and (\ref{Lambda}), we are
finally led to the two following, in principle equivalent, estimates
of the typical spherically symmetric specific angular momentum
profile,
\beq 
\sj(t) = \frac{J(t)}{M(t)}\Biggl\{
\frac{5}{2}-\frac{1}{\ra[M(t),t]} 
\left[\frac{\dot E}{2E(t)}-
\frac{\dot \lambda}{\lambda(t)}\right]\Biggr\} 
\label{sjdef}
\eeq
and 
\beqa
\sj(t) = \frac{J(t)}{M(t)}\Biggl\{
\frac{5}{3}-\frac{1}{\ra[M(t),t]}~~~~~~~~~~~~~~~~~ &\nonumber\\
\times\,\left[\frac{\der\ln(\delv\bar\rho)^{1/6}}{\der t}- 
\frac{\dot \lambda'}{\lambda'(t)}\right]
\Biggr\}\,,~~&
\label{sjdefp}
\eeqa
for $t(r)$ equal to the inverse of $R(t)$ and $J(t)$ the total
angular momentum accretion track, given by equations (\ref{lambda}) or
(\ref{lambdap}) for the appropriate values of $M(t)$, $E(t)$ and
$\lambda(t)$ or $\lambda'(t)$.

\begin{figure}
\centerline{\includegraphics[bb=18 190 592 650,clip,width=0.59\textwidth]
{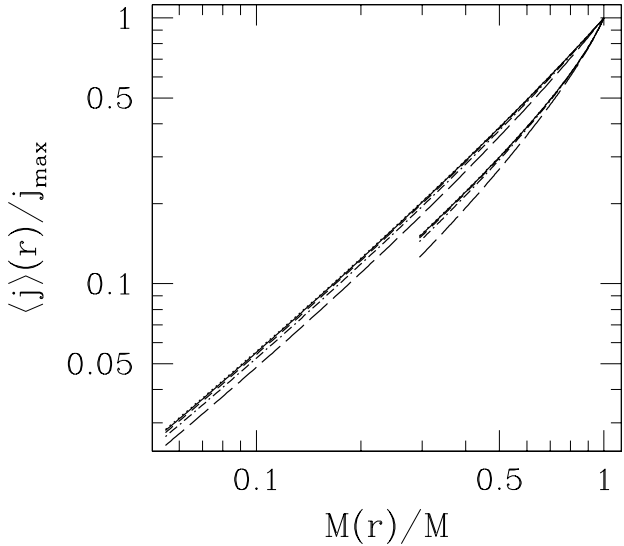}}
\caption{Same as Figures \ref{5} and \ref{3} but for the angular
momentum profile predicted using $\lambda$ for the extreme cases of
$10^{10}$ \modot (lower curves) and $10^{15}$ \modot (upper curves) in
order to show the degree of insensitivity to parameter $\eta$ for
solutions corresponding to different halo masses.}
\label{9}
\end{figure}

These two estimates of the angular momentum profile are shown in
Figure \ref{8}. As can be seen, the solutions derived from $\lambda$
and $\lambda'$ are slightly different from each other. This cannot be
due to the value of $\eta$ used to infer the solution from $\lambda$
because, as shown in Figure \ref{9}, the solutions appear to be
remarkably insensitive, in this occasion, to the value of that
parameter. We have also checked that the extrapolation of the density
profile beyond $R$ has a very small effect. Thus, it can only be due
to the time dependence adopted for $\lambda$ and $\lambda'$. In fact,
the information we have on the empirical behavior of the spin
parameters refers only to redshifts smaller than 2 (see Hetznecker and
Burkert 2006). For this reason, in Figure \ref{8}, the predicted
solutions are only drawn for radii involving $z\le 5$. 

Anyhow, both solutions show similar trends in agreement with the
results of numerical simulations. They are indeed power laws in mass
inside $r$ to a first approximation, particularly the solution
obtained from $\lambda$, with a mean best-fitting value of index $s$
(with the restriction $\sj(R)/j\maxi=1$) close to that found
empirically. More specifically, for the masses included in our figures
and a PS mass distribution, we obtain (within the radial range plotted
in Fig.~\ref{8} for each mass) $\bar s\sim 1.6$ and $\sim 1.7$ for the
$\lambda$ and $\lambda'$ cases, respectively. While for the mass
interval studied by B01b ($> 10^{12}$ \modotc) we obtain $\bar s\sim
1.3$ and $\sim 1.4$ to be compared with the value of $\sim 1.3$ quoted
by these authors. Furthermore, the slight bending upwards at large
radii, more apparent in the $\lambda'$ case although also present in
the $\lambda$ one as all the curves must approach unity at $r=R$, is
also observed in simulated halos (see the comments in B01b). Finally,
the tendency for less massive halos to have slightly steeper profiles
(in the mean all over the radial range analyzed) is also consistent
with the empirical results of B01b (see their Fig.~16).

\begin{figure}
\centerline{\includegraphics[bb=18 190 592 650,clip,width=0.59\textwidth]
{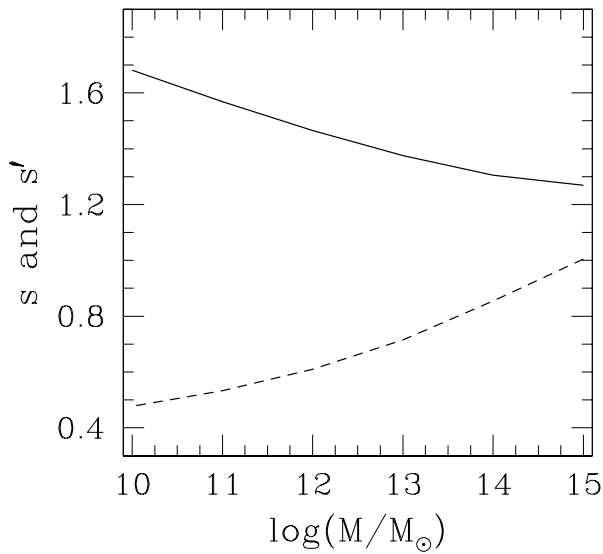}}
\caption{Mass dependence of the best fitting parameters $s$ and $s'$
in the power law and the modification of it proposed to fit the
specific angular momentum profile drawn from $\lambda$ (solid line) and
$\lambda'$ (dashed line), respectively.}
\label{11}
\end{figure}

As a power law does not seem to give a very good fit to the solutions
drawn from $\lambda'$ we have tried with another simple analytical
expression. As shown in Figure \ref{10}, the following slight
modification
\beq 
\frac{\sj(r)}{j\maxi} = 1-\left[1-\frac{M(r)}{M(R)}\right]^{s'},
\label{jnew}
\eeq
is already enough to improve notably the fits for all halo
masses. Note that the solutions drawn from $\lambda$ are however
better fit by the previous simple power law of index $s$. The
dependence on $M$ of that index $s$ and the new one $s'$ is
given in Figure \ref{11} for the solutions drawn from $\lambda$ and
$\lambda'$, respectively.

\section{SUMMARY AND CONCLUSIONS}\label{dis}

We consider the structural and kinematic properties of
realistic ellipsoidal, rotating, standard CDM halos in the
accretion-driven scenario. According to this scenario, halos adapt
dynamically to the amount of matter (carrying mass, energy and angular
momentum) that is accreted at any given moment. This is due to the
collisionless nature of CDM and the inside-out growth of these
structures during accretion phases. The typical accretion rate for
halos of a given mass at a given cosmic time depends only on the
cosmology, and thus, the typical radial profiles of relaxed dark
matter halos are fixed by cosmological parameters.

Specifically, we derive the typical spherically averaged profiles
corresponding to the density $\rho$, the velocity dispersion $\sigma$
and the phase-space density $\rho/\sigma^3$. We find that all these
predicted profiles are in good agreement with those found in numerical
simulations. In particular, $\rho$ is well fit, in 3D, by an Einasto
function and $\rho/\sigma^3$ by a power law, $Ar^{-\nu}$, with $\nu
\approx 1.9$ and $\log A\approx 1.4$.

We also derive the typical profile of the velocity anisotropy $\beta$.
We find that the anisotropy increases from something small (maybe
slightly negative) at the halo center to a positive value in the
outer region and with a typical logarithmic slope also in agreement
with numerical data. This is the first time that a non-zero anisotropy
has been derived from first principles. We find that the spherically
averaged phase-space density associated to the tangential (radial)
velocity dispersion, $\rho/\sigma\tang^3$ ($\rho/\sigma\rad^3$) is
also close to a power law with slope $\nu \approx 1.9$ although
somewhat different values of $A$.

We finally derive the typical spherically averaged profile of the
specific angular momentum $j$ from the empirical evolution of the
dimensionless halo spin parameter $\lambda$ or $\lambda'$. We find
that this profile scales, to a first approximation, as a power law in
inner mass, $j(r) \propto [M(r)]^s$, in agreement with numerical
simulations. The average logarithmic slope $s$ found using $\lambda$
coincides with the empirical value of 1.3, while that found using
$\lambda'$ (1.4) is also close to it. A more accurate analytical
expression is provided for the $j(r)$ profile derived from $\lambda'$.

We find that all the preceding profiles depend moderately on halo
mass. The dependence of the NFW and Einasto parameters fitting the
density profile in any given cosmology were already presented in
SMGH. Here we have focused on the mass dependence of the remaining
profiles. The behavior of any of these profiles found in observations
(for instance, the logarithmic slope of the density profile obtained
from X-rays or strong gravitational lensing) could in principle be
used, in a given cosmology, to estimate the total mass of the
system. Since the total mass most often is known, this also provides a
direct way of observationally testing the predictions of this
accretion driven scenario.

The anisotropy and velocity dispersion (or phase-space density)
profiles are derived using a practical phenomenological correction for
the effects of the complex process of shell-crossing on the total
energy of halos conserved since the epoch they were density
perturbations. This correction is achieved by adjusting the value of
one free parameter, $\eta$. We find that both the anisotropy profile
and the phase-space density profile are in good agreement with
numerical results for $\eta \approx 1.8$, the angular momentum profile
being instead insensitive to the value of this parameter.  

The fact that all our results are in fairly good agreement with those
drawn from numerical cosmological simulations leads support to the
idea that the structural and kinematic properties of dark matter
structures are determined by their dynamical adaption to cosmological
accretion.


\acknowledgments 

This work was supported by Spanish DGES grants AYA2003-07468-C03-01
and AYA2006-15492-C03-03. The Dark Cosmology Centre is funded by the
Danish National Research Foundation. We thank Andrei Doroshkevich for
fruitful discussions.


\appendix

\section{JEANS EQUATION AND SCALAR VIRIAL RELATION}
\label{App1}

\subsection{Exact Relations}
\label{genvir}

The internal dynamics of relaxed inside-out evolving halos should be 
well-described by a steady (true) phase-space density $f({\mathbf
r},{\mathbf v})$ satisfying the collisionless Boltzmann
equation. Writing this equation in spherical coordinates (see, e.g.,
equation [4p-2] of Binney \& Tremaine 1987), multiplying it by the
radial velocity $v\rad$, and integrating over velocity and solid
angle, one is led to the following first order differential equation
\beq
\frac{\der (\srho \sigma\rad^2)}{\der r}+ \frac{\srho(r)}{r} 
(3\sigma\rad^2(r) -
\sigma^2(r)) = -\frac{1}{4\pi}\isph{\rho\,\derpr\Phi}\,,
\label{Jeq1}
\eeq
where $\sigma$ stands for the non-centered velocity dispersion in the
shell at $r$,
\beq
\sigma^2(r)=\frac{1}{4\pi\srho}\isph{\int\der^3 v\,v^2\,f}=
\sigma^2\rad(r)+\sigma^2_\theta(r)+ \sigma^2_\varphi(r)\,,
\eeq
$\srho$ for the spherically averaged density profile of the halo
given by equation (\ref{avden}), with the local density distribution satisfying
the relation
\beq \rho\sphc = \int\der^3 v\,f({\mathbf r},{\mathbf v})\,,  
\eeq 
and $\derpr$ for the radial partial derivative.  To derive equation
(\ref{Jeq1}), it is only needed such conventional assumptions as the
continuity, in real space, of the local density and mean velocities
and the fact that $f$ vanishes for large velocities.  In expression
(\ref{Jeq1}), we have taken into account that the effects of a
non-null cosmological constant are negligible on halo scales.

Splitting the local density and gravitational potential as
\begin{eqnarray}
\rho\sphc &=&\srho(r)+\delta\rho\sphc
\label{split1}\\
\Phi\sphc &=&\spot(r)+\delta\Phi\sphc\,,
\label{split2}
\end{eqnarray}
and taking into account that the spherically averaged gravitational potential,
\beq
\spot(r) = \frac{1}{4\pi}\isph{\Phi\sphc} ,
\label{spot}
\eeq
satisfies, by the Gauss theorem, the usual relation for spherically
symmetric systems
\beq
\frac{\der\spot(r)}{\der r} = \frac{G M(r)}{r^2} ,
\label{gauss}
\eeq
equation (\ref{Jeq1}) adopts the form
\beq
\frac{\der (\srho \sigma\rad^2)}{\der r}+ \frac{\srho}{r} (3\sigma\rad^2 -
\sigma^2) = -\srho\frac{GM(r)}{r^2} - \frac{1}{4\pi}\isph{\delta\rho\,
\derpr(\delta\Phi)}\,.
\label{exJeq}
\eeq

Except for the second term on the right, equation (\ref{exJeq}) looks
exactly as the classical Jeans equation for spherically symmetric
systems with anisotropic velocity tensor. Thus, multiplying (\ref{exJeq})
by $4\pi r^3\der r$ and integrating over $r$, the same steps
leading to the scalar virial relation for a spherically symmetric
system now lead to
\beq
4\pi R^3 P(R) - 2K = -\int_0^R \der M(r)\frac{GM(r)}{r} - \int_0^R
\der r\, r^3 \isph{\delta\rho\, \derpr (\delta\Phi)} \,,
\label{vir1}
\eeq
where $\der M(r)=4\pi r^2\srho(r)\der r$ is the mass of the elementary
spherical shell of radius $r$,
\beq
P(R)=\srho (R)\sigma\rad^2(R)
\label{pressure}
\eeq
is the spherically averaged radial boundary pressure, and 
\beq
K=\frac{1}{2}\int_0^R \der M(r)\sigma^2(r)
\label{kinexac}
\eeq
is the kinetic energy within $R$.

On the other hand, the total potential energy is
\beq W=\frac{1}{2}\int_0^R \der r\,
r^2\isph{\rho\sphc\,\Phi\sphc}\,,
\eeq
which, from equations (\ref{split1})--(\ref{split2}), can be written as
\beq
W=2\pi\int_0^R \der r\,r^2\srho(r)\,\spot(r) + \frac{1}{2}\int_0^R\der
r\,r^2\isph{\delta\rho\,\delta\Phi}\,.
\label{potexac}
\eeq
Provided the central asymptotic logarithmic slope of $\srho$ is
greater than $-5/2$, fixing the origin of the potential $\Phi$ so to
have the boundary condition 
\beq
\spot(R) = -\frac{GM}{R}\,,
\label{bound}
\eeq
for the differential equation (\ref{gauss}) (this is essentially
equivalent to consider the potential origin at infinity for the system
truncated at $R$) and integrating by parts (two consecutive times) the
first term on the right of equation (\ref{potexac}), the potential
energy takes the form
\beq
W=-\int_0^R\der M(r)\,\frac{GM(r)}{r}+ \frac{1}{2}\int_0^R\der
r\,r^2\isph{\delta\rho\,\delta\Phi}\,.
\label{potsph}
\eeq
Therefore, by subtracting $2W$ on both sides of equation (\ref{vir1})
we arrive to the virial relation
\beq
4\pi R^3 P(R) - 2E = \int_0^R \der M(r)\,\frac{GM(r)}{r}- \int_0^R\!\der r\,
r^2\!\isph{\delta\rho\,\left[\delta\Phi+r\derpr (\delta\Phi)\right]}\,.
\label{virial}
\eeq
where the total energy $E=K+W$ is given by
\beq
E = \int_0^R \der M(r)\,\left[\frac{\sigma^2}{2}-\frac{GM(r)}{r}\right]+
\frac{1}{2}\int_0^R\der r\,r^2\isph{\delta\rho\,\delta\Phi}\,.
\label{ener}
\eeq

\subsection{Approximate Relations}\label{aproxvir}

Although halos exhibit substantial triaxiality (the average minor to
major axial ratio takes a value between $0.6$ and $0.7$; Bullock 2002,
Kasun \& Evrard 2005, Bailin \& Steinmetz 2005, Libeskind et
al. 2005), their mass distribution is far from flattened and the
isopotential surfaces are more spherical (cf. Binney \& Tremaine
1987). We therefore have
\begin{eqnarray}
\frac{|\delta\rho|}{\srho} &<& 1\label{ine1}\\
\frac{|\delta\Phi|}{|\spot|}&\ll& 1\,.\label{ine2}
\end{eqnarray}
Moreover, since the isopotential surfaces approach to spheres as $r$
increases, the rms value of $\delta\Phi$ in the spherical shell at $r$,
\beq
\rms(r) = \left[\frac{1}{4\pi}\isph{(\delta\Phi)^2}\right]^{1/2}\,,
\label{rmsphi}
\eeq
is a positive monotonously decreasing function of $r$ satisfying the inequality
\beq
\frac{\rms}{|\spot|} \ll 1\,.
\label{ineq2}
\eeq
Then, the fact that $|\spot|$ is also a monotonous positive
decreasing function of $r$ (see eqs.~[\ref{bound}] and [\ref{gauss}])
implies
\beq
\left|\frac{\der \rms(r)}{\der r}\right| \ll \frac{GM(r)}{r^2}\,.
\label{cond1}
\eeq
To see it, consider the proportionality
\beq
\rms(R)=\epsilon |\spot(R)|
\label{prop1}
\eeq
with $\epsilon$ much smaller than unity, as implied by the
inequality (\ref{ineq2}). If we move to the halo center, the positive
increments $\Delta\rms\equiv \rms(0)-\rms(R)$ and $\Delta|\spot|\equiv
|\spot(0)|-|\spot(R)|$ cannot satisfy the $\epsilon$-order relation
\beq
\Delta\rms\sim \epsilon^{\mu} \Delta |\spot|\,,
\label{prop4}
\eeq
with $\mu$ any {\em negative integer or null} number, because one would then have
\beq
\frac{\rms(0)}{|\spot(0)|}=\frac{\rms(R)+\Delta\rms}{|\spot(R)|+\Delta |\spot|}
=\frac{\epsilon+\frac{\Delta \rms}{|\spot(R)|}}
{1+\frac{\Delta |\spot|}{|\spot(R)|}}\sim 
\frac{\epsilon+\epsilon^{\mu}\frac{\Delta |\spot|}{|\spot(R)|}}
{1+\frac{\Delta |\spot|}{|\spot(R)|}} \ga 1\,,
\label{prop5}
\eeq
with the last inequality arising from the condition (\ref{prop4}) and
the fact that $\Delta |\spot|/|\spot(R)|$ is of order unity (see
eqs.~[\ref{gauss}] and [\ref{bound}]), which contradicts the condition
(\ref{ineq2}). Thus, they must necessarily satisfy instead
\beq
\Delta\rms\sim \epsilon^\mu \Delta |\spot|
\label{prop2}
\eeq
with $\mu$ some {\em positive integer} number, in which case the relations
\beq
\frac{\der \rms}{\der r}\sim \frac{\Delta\rms}{R} \sim \epsilon^\mu
\frac{\Delta |\spot|}{R}\sim \epsilon^\mu \frac{\der |\spot|}{\der r}\,,
\label{prop3}
\eeq
validate the inequality (\ref{cond1}).

For fixed values of $\theta$ and $\phi$, $|\delta\Phi|$ is a
continuous function of $r$ with typical amplitude equal to $\rms$
except in those directions where it vanishes (this necessarily
happens in specific radial directions owing to the homologous
character of triaxial halos). We therefore have
\beq
|\derpr\delta\Phi|=|\partial\rad |\delta\Phi||\sim 
\left|\frac{\der \rms}{\der r}\right|\,,
\label{cond2}
\eeq
so equations (\ref{ine1}) and (\ref{cond1}) lead to
\beq
|\delta\rho|\,|\derpr\delta\Phi| \ll \srho(r)\,\frac{\,GM(r)}{r^2}\,,
\label{cond3}
\eeq
the latter inequality also holding now over those radial directions
where $\delta \Phi$ vanishes and, hence, $\derpr\delta\Phi$ is
identically null.

The inequality (\ref{cond3}) allows one to write equation
(\ref{exJeq}) in the approximate form
\beq
\frac{\der (\srho \sigma\rad^2)}{\der r}+ \frac{\srho(r)}{r}\,[3\sigma^2\rad(r) -
\sigma^2(r)] \approx -\srho(r)\,\frac{GM(r)}{r^2} \label{exJeq2}
\eeq
This expression coincides with the Jeans equation for spherically
symmetric systems in terms of the non-centered velocity dispersion
profiles. Therefore, proceeding in the same way as
leading to equation (\ref{virial}) we obtain approximately the usual
scalar virial relation for spherically symmetric systems
\beq
4\pi R^3 P(R) - 2E \approx \int_0^R \der M(r)\frac{GM(r)}{r}\,,
\label{virapr}
\eeq
where $P(R)$ is given by equation (\ref{pressure}) and the total
energy $E$ takes, from equations (\ref{kinexac}) and (\ref{potsph}) and
the inequalities (\ref{ine1})-(\ref{ine2}), approximately the same
form as in the spherically symmetric case,
\beq E=K+W\approx 4\pi\int_0^R \der r\, r^2\,\srho(r)
\left[\frac{\sigma^2(r)}{2}- \frac{GM(r)}{r}\right]\,.
\label{etot}
\eeq
%


\end{document}